\documentstyle[preprint,tighten,eqsecnum,aps,floats,psfig,epsfig]{revtex}

\newcommand{\fd}{fluc\-tu\-a\-tion-dis\-si\-pa\-tion }
\newcommand{\beq}{\begin{equation}}
\newcommand{\beqa}{\begin{eqnarray}}
\newcommand{\eeq}{\end{equation}}
\newcommand{\eeqa}{\end{eqnarray}}

\newcommand{\dpar}{\partial}
\renewcommand{\d}{{\rm d}}
\newcommand{\Li}{{\rm Li_2}}

\begin{document}
\draft

\title{
Two-loop Critical Fluctuation-Dissipation Ratio \\ 
for the Relaxational Dynamics of the $O(N)$ \\
Landau-Ginzburg Hamiltonian
}
\author{Pasquale Calabrese and Andrea Gambassi}
\address{Scuola Normale Superiore and INFN,
Piazza dei Cavalieri 7, I-56126 Pisa, Italy. 
\\
{\bf e-mail: \rm
{\tt calabres@df.unipi.it},
{\tt andrea.gambassi@sns.it}
}
}

\date{\today}

\maketitle

\begin{abstract}
The off-equilibrium purely dissipative dynamics (Model A) of the $O(N)$ 
vector model is considered at criticality in an 
$\epsilon =  4- d > 0$ expansion up to $O(\epsilon^2)$.
The scaling behavior of two-time
response and correlation functions at zero momentum, 
the associated universal scaling functions 
and the nontrivial limit of the fluctuation-dissipation ratio 
are determined  in the aging regime.    
\end{abstract}

\pacs{PACS Numbers: 64.60.Ht, 05.40.-a, 75.40.Gb, 05.70.Jk}


\section{Introduction}
\label{intr}

In recent years many efforts have been made in order to understand
the off-equilibrium aspects of the dynamics of statistical systems.
A variety of novel dynamical behaviors emerges when some kind 
of randomness is present in the system. Among them, one of the most 
striking is that of {\it aging} (see Ref.~\cite{review} and references 
therein). 
It has been pointed out~\cite{ckp-94} that they
 could also emerge in nondisordered systems if
slow-relaxing modes are present. 
This naturally happens when
the system undergoes a second-order phase transition at some 
critical temperature $T_c$. 
Indeed, 
consider a ferromagnetic model in a disordered state and
quench it to a given temperature $T\geq T_c$~\cite{footnote0}
at time $t=0$.
During the relaxation a small external field $h$ is applied at ${\bf x}=0$
after a waiting time $s$. At time $t$, the order parameter response to $h$ is
given by the response
function $R_{\bf x} (t,s)=\delta\langle \phi_{\bf x} (t)\rangle/ \delta h (s)$,
where $\phi$ is the order parameter and $\langle \cdot \rangle$
stands for the mean over the stochastic dynamics.
Correlations of order parameter fluctuations are interesting dynamical
quantities as well. In the following we will focus on the two-time one, 
given by
$C_{\bf x} (t,s)=\langle \phi_{\bf x} (t)\phi_{\bf 0} (s)\rangle$.
The time evolution of the model we are considering is characterized 
by two different regimes:
a transient one with off-equilibrium evolution, for $t<t_R$, and a
stationary equilibrium evolution for $t>t_R$, where $t_R$ is the relaxation
time.
In the former a dependence of the system behavior on initial condition
is expected, while in the latter time homogeneity and time reversal
symmetry~(at least in the absence of external fields) are recovered; as a
consequence we expect that for $t_R\ll s,t,\; R_{\bf x} (t,s)=R_{\bf x}^{\rm eq}(t-s),\; C_{\bf x} (t,s)=C_{\bf x}^{\rm eq}(t-s)$ where $R_{\bf x}^{\rm eq}$
and $C_{\bf x}^{\rm eq}$ are determined by the ``equilibrium'' dynamics of
the system, with a characteristic time scale diverging at
criticality~(critical slowing down).
Moreover the \fd theorem states that
\beq
R_{\bf x}^{\rm eq}(\tau)=-{1\over T} {\d C_{\bf x}^{\rm eq}(\tau)\over\d\tau}.
\eeq
 
If the system does not reach the equilibrium, all the previous functions will
depend both on $s$~(the ``age'' of the system) and $t$.
This behavior is usually referred to as aging and was
first predicted for spin glass systems~\cite{review,ck-94}.
The
\fd ratio~(FDR) \cite{ckp-94,ck-94}:
\beq
X_{\bf x}(t,s)=\frac{T\, R_{\bf x}(t,s)}{\dpar_s C_{\bf x}(t,s)} \; ,
\label{dx}
\eeq
is usually introduced to measure the distance from the equilibrium 
of an aging system evolving at a fixed temperature $T$.
A non trivial value for this ratio is also experimentally observed 
in some glassy systems \cite{exp}.

In recent years several works \cite{review,ckp-94,ck-94,nb-90,cd-95,ckp-97,barrat,bbk-99,fmpp-98,zkh-00,cst-01,clz} have been devoted to the study of the FDR
for systems exhibiting domain growth~\cite{bray},
and for aging systems such as glasses and spin glasses. 
$X_{\bf x}(t,s)$ turns out to be a nontrivial function of $t$ and $s$, in the 
low-temperature phase of all these systems. 
In particular, analytical and numerical studies indicate that the limit
\beq
X^\infty=\lim_{s\to\infty}\lim_{t\to\infty}X_{{\bf x}=0}(t,s),
\label{xinfdef}
\eeq
vanishes throughout the low-temperature phase both for spin glasses and simple
ferromagnetic systems~\cite{ckp-97,barrat,bbk-99,zkh-00,cst-01}.

Only recently \cite{ckp-94,lz-00,gl-00,bhs-01,gl-000,gl-02b,cg-02} 
attention has been paid to the FDR, for nonequilibrium, nondisordered, 
and unfrustrated systems  at criticality.
From general scaling arguments one would expect that the critical
response function scales as \cite{gl-000,gl-02b,cg-02,jss-89,j-92}:
\beq
R_{\bf x=0}(t,s)= {\cal A}_R (t-s)^{a-d/z} (t/s)^\theta {\cal F}_R(s/t) \, ,
\label{Rscalform}
\eeq
where $a=(2-\eta-z)/z$ and 
$\theta$ is the initial-slip exponent of the response function, related
to the initial-slip exponent of the magnetization $\theta '$ 
and to the autocorrelation exponent $\lambda_c$~\cite{huse-89}
by the relation \cite{jss-89}
\beq
\theta' =\theta +z^{-1}(2-z-\eta)=z^{-1}(d-\lambda_c).
\eeq
In recent works the notion of local scale invariance has been introduced
as an extension of anisotropic or dynamical scaling (see \cite{henkel-02}
and references therein). Assuming that the response function transforms
covariantly under the constructed group of local transformations, it has 
been argued \cite{hpgl-01} that ${\cal F}_R(s/t)=1$.
Under the same assumption, the full spatial dependence has been also predicted
\cite{henkel-02}
\beq
R_{\bf x}(t,s)=R_{\bf x=0}(t,s) \Phi (|{\bf x}|/(t-s)^{1/z}),
\label{henkelpred}
\eeq
where $\Phi(u)$ is a function whose convergent series expansion is explicitly
 known \cite{henkel-02}.
For the correlation function and its derivative no analogous prediction
exists.
One can only expects from general Renormalization Group (RG) arguments that 
\cite{gl-000,gl-02b,cg-02,jss-89,j-92}
\beqa
C_{\bf x=0}(t,s)&=&{\cal A}_C (t-s)^{a+1-d/z} (t/s)^{\theta-1} {\cal F}_C (s/t)\, ,
\label{Cscalform} \\
\dpar_s C_{\bf x=0}(t,s)&=&{\cal A}_{\dpar C}(t-s)^{a-d/z} (t/s)^\theta {\cal F}_{\dpar C} (s/t)\, ,
\label{dCscalform} 
\eeqa
with the same $\theta$ and $a$ as in Eq.~(\ref{Rscalform}).
The functions ${\cal F}_C(v)$, ${\cal F}_{\dpar C}(v)$ 
and ${\cal F}_R(v)$ are universal and 
defined in such a way 
${\cal F}_C(0)={\cal F}_{\dpar C}(0) = {\cal F}_R(0) = 1$, while the constants
${\cal A}_C$, ${\cal A}_{\dpar C}$ and ${\cal A}_R$ are 
nonuniversal amplitudes.

From the scaling laws of above, it has been argued 
that $X^\infty$ is a {\it new universal} quantity 
associated with the given nonequilibrium dynamics~\cite{gl-000,gl-00,gl-02b},
 and, as such, it should attract the same 
interest as critical exponents. Given this universality, it is worthwhile 
to compute $X^\infty$ for those mesoscopic models of
dynamics which have the same critical behavior of
some lattice models considered so far in the literature. 

Correlation and response functions were exactly computed in the simple cases
of a Random Walk, a free Gaussian field, and a two-dimensional $XY$ model at 
zero temperature and the value $X^\infty=1/2$ 
was found \cite{ckp-94}.
The analysis of the $d$-dimensional spherical model gave $X^\infty=1-2/d$
\cite{gl-00}, while $X^\infty=1/2$ for the one-dimensional Ising-Glauber 
chain \cite{lz-00,gl-000}.
Monte Carlo simulations have been done for the 
two- and three-dimensional Ising model \cite{gl-00}, finding
$X^\infty=0.26(1)$ and $X^\infty\simeq 0.40$ 
respectively.
The effect of long-range correlations in the initial configuration has been  
also analyzed for the $d$-dimensional spherical model \cite{ph-02}.

Only in a recent work~\cite{cg-02} field-theoretical methods have been
applied to determine the FDR and the scaling forms of the response and 
correlation functions up to the first order in an $\epsilon$ expansion,
for the purely relaxational dynamics of the $O(N)$ Ginzburg-Landau model. 
This field-theoretical model has the same symmetries 
(and thus the same universal properties) as a wide class of spin 
systems on the lattice with short range interactions 
(see~\cite{ZJ-book}, or~\cite{PV-r} for a recent review). 

In~\cite{cg-02} the following  quantity, related to the FDR, 
was introduced  in momentum space
\beq
{\cal X}_{\bf q}(t,s)=
{\Omega R_{\bf q}(t,s)\over \dpar_s C_{\bf q}(t,s)},\label{Xq}
\eeq
where $R_{\bf q}(t,s)$ and $C_{\bf q}(t,s)$ are the Fourier transforms~(with
respect to ${\bf x}$) of $R_{\bf x}(t,s)$ and $C_{\bf x}(t,s)$ respectively.
It was argued that the zero-momentum limit 
\beq
{\cal X}_{\bf q=0}^{\infty}=
\lim_{s\to\infty}\lim_{t\to\infty} {\cal X}_{\bf q=0}(t,s)
\label{Xq2}
\eeq
is equal to the same limit of the FDR~(\ref{xinfdef}) for ${\bf x=0}$, i.e. 
${\cal X}_{\bf q=0}^\infty=X^\infty$ to all orders~\cite{cg-02}. 
This fact allows an easier perturbative computation in momentum space 
of the new universal quantity $X^\infty$.

The extension of these results to two-loop order is very important not only
from a quantitative point of view. 
In fact, in the past, when new scaling relations have been proposed,
several times they resisted to the test of the first order in the $\epsilon$ 
expansion, but not to higher order calculations.
Classical examples may be found in 
the context of surface criticality (see e.g. the review \cite{diehl}, p. 116,
and references therein) and in the case of anisotropic scaling at  
Lifshitz points \cite{lif}.
This is a further reason to present here the second order computation of
the scaling form for the zero-momentum response function and 
the FDR for the purely dissipative relaxation of the $O(N)$ model.

The paper is organized as follows. 
In Sec. \ref{sec2} we briefly introduce the model. 
In Sec. \ref{sec3} we evaluate the zero-momentum response function 
$R_{\bf q=0}(t,s)$ and in particular we derive its scaling form. 
In Sec. \ref{sec4} we compute the FDR up to the second order in $\epsilon$ and
we derive a scaling form for $\dpar_s C_{\bf q=0}(t,s)$.
In Sec. \ref{sec5} we summarize and comment our results and discuss some 
points that need further investigation.
In the Appendices \ref{a1} and \ref{a2} we give all the details to
compute the zero-momentum Feynman integrals. 

\section{The model}
\label{sec2}

The time evolution of an $N$-component field $\varphi({\bf x},t)$ under a
purely dissipative dynamics~(Model A of Ref.~\cite{HH}) is 
described by the stochastic Langevin equation
\beq
\label{lang}
\dpar_t \varphi ({\bf x},t)=-\Omega 
\frac{\delta \cal{H}[\varphi]}{\delta \varphi({\bf x},t)}+\xi({\bf x},t) \; ,
\eeq
where $\Omega$ is the kinetic coefficient, 
$\xi({\bf x},t)$ a zero-mean stochastic Gaussian noise with 
\beq
\langle \xi_i({\bf x},t) \xi_j({\bf x}',t')\rangle= 2 \Omega \, \delta({\bf x}-{\bf x}') \delta (t-t')\delta_{ij},
\eeq
and $\cal{H}[\varphi]$ is the static Hamiltonian. 
It may be assumed, near the critical point,
of the Landau-Ginzburg form
\beq
{\cal H}[\varphi] = \int \d^d x \left[
\frac{1}{2} (\nabla \varphi )^2 + \frac{1}{2} r_0 \varphi^2
+\frac{1}{4!} g_0 \varphi^4 \right] .\label{lgw}
\eeq
  
Instead of solving the Langevin equation for $\varphi(\xi)$ and then 
averaging over the noise distribution, the equilibrium  correlation and 
response functions can be directly obtained by means of the 
field-theoretical action \cite{ZJ-book,bjw-76} 
\beq
S[\varphi,\tilde{\varphi}]= \int \d t \int \d^dx 
\left[\tilde{\varphi} \frac{\partial\varphi}{\partial t}+
\Omega \tilde{\varphi} \frac{\delta \mathcal{H}[\varphi]}{\delta \varphi}-
\tilde{\varphi} \Omega \tilde{\varphi}\right].\label{mrsh}
\eeq
Here $\tilde{\varphi}({\bf x},t)$ is an auxiliary field, conjugated to 
the external field $h$ in such a way that
${\cal H}[\varphi,h] = {\cal H}[\varphi] - \int d^d x h \, \varphi$.
As a consequence, the linear response to the field $h$ of a generic observable
${\cal O}$ is given by
\beq
{\delta \langle {\cal O} \rangle \over \delta h({\bf x},s)} = 
\Omega \langle \tilde\varphi({\bf x},s){\cal O}\rangle \ ,
\eeq
for this reason   $\tilde{\varphi}({\bf x},t)$ is termed response field.

The effect of a macroscopic initial condition 
$\varphi_0({\bf x})=\varphi({\bf x},t=0)$ may be taken into account by 
averaging over the initial configuration
with a weight $e^{-H_0[\varphi_0]}$ where, for example, 
\beq
H_0[\varphi_0]=\int\! \d^d x\, \frac{\tau_0}{2}(\varphi_0({\bf x})-a({\bf x}))^2,
\eeq
that specifies an initial state $a({\bf x})$ with gaussian short-range 
correlations proportional to $\tau_0^{-1}$. 
Any addition of anharmonic terms in $H_0[\varphi_0]$ is not expected to be 
relevant as long as the harmonic term is there~(as in the case when the initial
state is in the high temperature phase).
Instead, an initial condition with long-range correlations
may lead to a different universality class, as e.g. shown for the 
$d$-dimensional spherical model with non-conservative dynamics \cite{ph-02}.

Following standard methods \cite{ZJ-book,bjw-76},
the response and correlation functions may
be obtained by a 
perturbative expansion of the functional weight
$e^{-(S[\varphi,\tilde{\varphi}]+H_0[\varphi_0])}$ in terms of the
coupling constant $g_0$ (appearing in the vertex 
$g_0 \varphi^3 \tilde{\varphi} /3!$).
The propagators~(Gaussian two-point functions of the fields $\varphi$ and 
$\tilde{\varphi}$) of the resulting theory are \cite{jss-89} 
\beqa
\langle \tilde{\varphi_i}({\bf q},s) \varphi_j(-{\bf q},t) \rangle_0 =& 
\delta_{ij} R^0_q(t,s)=&\delta_{ij} \,\theta(t-s) G(t-s),\label{Rgaux}\\
\langle \varphi_i({\bf q},s) \varphi_j(-{\bf q},t) \rangle_0 =&
\delta_{ij} C^0_q(t,s)=& {\delta_{ij} \over q^2+r_0}\left[ G(|t-s|)+\left(\frac{r_0 +q^2}{\tau_0}-1
\right) G(t+s)\right], \label{Cgaux}
\eeqa
where
\beq
G(t)=\displaystyle{e^{-\Omega (q^2+r_0) t}} \label{GG}.
\eeq
The response function Eq.~(\ref{Rgaux}) is the same as in 
equilibrium.
Eq.~(\ref{Cgaux}), instead, reduces to the equilibrium form when both
times $t$ and $s$ go to infinity and $\tau = t - s$ is kept fixed.
In the following we will assume the \^Ito prescription 
(see~\cite{j-92,lrt-79}, \cite{ZJ-book} and references therein)
to deal with the 
ambiguities that arise in formal manipulations of stochastic equations. 
Consequently, all the diagrams with 
self-loops of response propagator has to be omitted in the computation. 
This ensures that causality holds in the perturbative 
expansion~\cite{jss-89,j-92,bjw-76}.
From the technical point of view, the breaking of time homogeneity makes 
the renormalization procedure in terms of one-particle 
irreducible correlation functions less straightforward than in 
standard cases~(see Ref. \cite{diehl,jss-89,j-92}).
Thus the computations will be done in terms of connected functions.

From the expressions above, we can compute the FDR for the Gaussian model
\cite{ckp-94,cg-02}: 
\beq
\displaystyle
{\cal X}_{\bf q}^0(t,s)=\left({\dpar_s C^0_q \over \Omega R^0_q}\right)^{-1}=
\left(1+e^{-2 \Omega (q^2+r_0) s}
+\Omega q^2 \tau_0^{-1} e^{-2 \Omega (q^2+r_0) s} \right)^{-1}.
\eeq
When the model is not at its critical point, i.e. $r_0 \propto T-T_c \neq 0$, 
the limit of this ratio for 
$s\rightarrow \infty$ is $1$ for all values of $q$, according to 
the idea that in the high-temperature phase all modes have a 
finite equilibration time. In this case equilibrium is approached 
exponentially fast in time and as a consequence the \fd theorem applies.
For the critical model, i.e. $r_0=0$, if $q\neq 0$
then the limit ratio is again equal to one, whereas for $q=0$ we have
${\cal X}_{q=0}^0(t,s)=1/2$. We can argue that, in the Gaussian model, 
the only mode characterized by aging, i.e. that ``does not relax'' to the 
equilibrium, is the zero mode in the critical limit. 

\section{Two-loop response function}
\label{sec3}

\begin{figure}[t]
\centerline{\epsfig{file=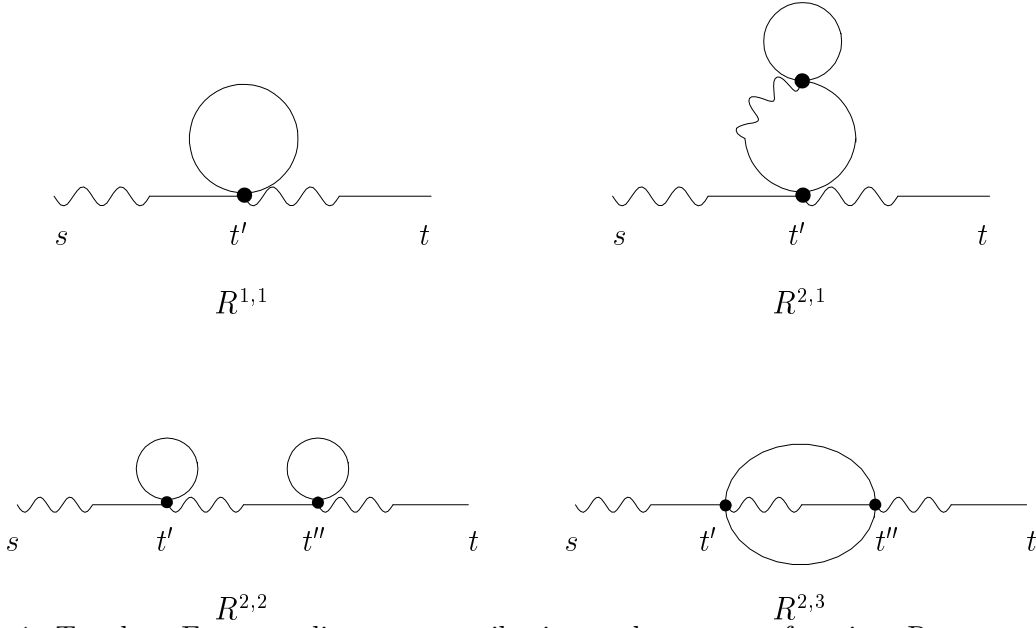}} 
\caption{Two-loop Feynman diagrams contributing to the response function. 
Response propagators are drawn as wavy-normal lines, whereas correlators are 
normal lines. A wavy line is attached to the response field 
and a normal one to the order parameter.}\label{figR}
\end{figure}

In this Section we compute, up to the second order in a loop expansion,
the critical nonequilibrium  response function at zero external momentum for 
the model described in the previous Section.
We use here the method of renormalized field theory in dimensional 
regularization with minimal subtraction of dimensional poles.
Up to the second order in perturbation theory there are  
four connected  Feynman 
diagrams (without self-loops of response propagator) that contribute 
to the response function. 
They are depicted in Figure~\ref{figR}.
In terms of these diagrams and as a function of the bare couplings and fields 
(denoted in the following with $\varphi_B$, $\tilde\varphi_B$),
the zero-momentum bare response function $R_B(t,s)$ is given by
\beqa
R_B(t,s)&=& R^0_{q=0}(t,s)- {N+2 \over 6} g_0 R^{1,1} \nonumber\\
&&+g_0^2 \left[ \left({N+2 \over 6}\right)^2 R^{2,1}+ {(N+2)^2\over18} R^{2,2}
+{N+2 \over 6} R^{2,3} \right] +O(g_0^3) \;.
\eeqa

In the following we assume $t>s$ for simplicity. We also fix 
$\tau_0^{-1}=0$, since  $\tau_0^{-1}$ is an irrelevant variable 
(in the RG sense) and thus it affects only the corrections to the leading 
scaling behavior \cite{jss-89,j-92}.
Using the results reported in the Appendix \ref{a1}, we get
\beqa
R_B(t,s)&=&1+ \tilde{g}_0 {N+2 \over 24} \left\{\log {t\over s}+ 
{\epsilon \over2} \left[\left(\gamma_E +\log2+\log t \right) \log  {t\over s}- 
{1\over2}\log^2 {t\over s} \right]  \right\}  +\nonumber\\ 
&&{\tilde{g}_0^2\over144} \left\{{(N+2)^2 \over 8} \log^2 {t\over s} + 
(N+2)^2 \left[-\left( {1\over \epsilon}+ \log2 +\gamma_E+ \log t  \right)
\log {t\over s} + {1\over2}\log^2 {t\over s}\right] \right\}\nonumber\\
&&-\tilde{g}_0^2{N+2\over24}
\left[ {1\over \epsilon}\left(\log {4\over3}+\log{t\over s}\right) 
+\log {t\over s}\left({1\over2}+\log t +\gamma_E\right) 
-{1\over2}\log^2 {t\over s}
\right.\nonumber\\ &&\left.          
+(\log(t-s)+\gamma_E)\log {4\over3} -{f(s/t)\over4}
 \right] 
+O(g_0^3,g_0^2 \epsilon, g_0 \epsilon^2)\, ,
\eeqa
where $\tilde{g}_0=N_d g_0$, $N_d=2/((4\pi)^{d/2}\Gamma(d/2))$ and 
$f(v)$ is a regular function defined in Eq. (\ref{fvd}). 
To lighten the notations we set $\Omega = 1$ in the previous equations.
The dependence on $\Omega$ of final formulas may be simply obtained 
by $t\mapsto\Omega t$, where $t$ is the generic time variable. 

\begin{figure}[t]
\centerline{\epsfig{width=8truecm,file=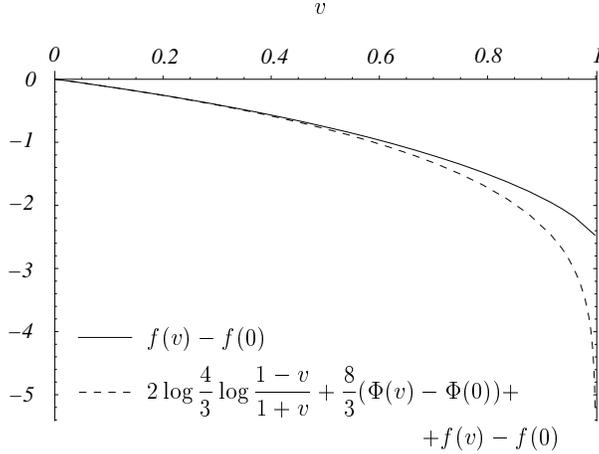}} 
\caption{Plot of the two-loop contribution 
to the universal functions $F_R(v)$ (see Eq.~(\ref{eqGR})) and $F_{\partial C}(v)$ (see Eq.~(\ref{eqGdC})). } \label{fv}
\end{figure}

In order to cancel out the dimensional poles appearing in this function, 
we have to renormalize the coupling constant according to \cite{ZJ-book}
\beq
\tilde{g}_0=\left(1+{N+8\over 6}{\tilde{g}\over \epsilon}\right)\tilde{g}+
O(\tilde{g}^2),
\label{rencou}
\eeq
and the fields $\varphi$ and $\tilde{\varphi}$ via the relations 
\cite{bjw-76} 
$\varphi_B = Z_{\varphi}^{1/2} \varphi $, $\tilde{\varphi}_B= Z_{\tilde{\varphi}}^{1/2}\tilde{\varphi}$, 
so that
\beq
R(t,s)= (Z_{\varphi} Z_{\tilde{\varphi}})^{-1/2} R_B(t,s)=
\left[1+{N+2\over24}\log{4\over3}{\tilde{g}^2\over\epsilon}+
O(\tilde{g}^3)\right] R_B(t,s).
\label{renfie}
\eeq
After this renormalization, $R(t,s)$ is a regular function of dimensionality 
also for $\epsilon\rightarrow 0$.
The critical response function is now obtained by fixing $\tilde{g}$ at 
its fixed point value \cite{ZJ-book}
\beq
\tilde{g}^*={6\,\epsilon\over N+8}\left[1+{3(3N+14)\over(N+8)^2}\epsilon\right]+O(\epsilon^3)\, ,\label{gstar}
\eeq
leading to
\beqa
R(t,s)&=&1+\epsilon {N+2\over 4(N+8)}\log {t\over s} +
{\epsilon^2 \over4}\left[
{6(N+2)\over (N+8)^2}\left({N+3\over N+8}+\log2\right) \log{t\over s}+
{(N+2)^2\over 8(N+8)^2}\log^2 {t\over s}\right.\nonumber\\
&&\left. -{6(N+2)\over (N+8)^2}\log {4\over3}\log(t-s)+
{3(N+2)\over 2(N+8)^2} \left( f(s/t)-4\gamma_E \log{4\over3} \right)\right] 
+ O(\epsilon^3)\; .
\label{R2loop}
\eeqa
Note that the nonscaling terms, like $\log t \log t/s$ 
(appearing, for example, in $R^{2,3}$, see Eq.~(\ref{eqR23})), cancel 
each other out when the coupling constant is set equal to its 
fixed point value. Eq.~(\ref{R2loop}) agrees
with the expected scaling form in momentum space 
(analogous to that in real space, Eq.~(\ref{Rscalform}))
\beq
R(t,s)=A_R (t-s)^a (t/s)^\theta F_R(s/t)\, ,
\eeq
with the well-known exponents \cite{jss-89,j-92,ZJ-book}
\beqa
\theta&=&{N+2\over N+8}{\epsilon \over4}\left[1+{6\epsilon\over N+8}
\left({N+3\over N+8}+\log2\right)\right]+O(\epsilon^3)\, , \\
a&=&{2-\eta-z\over z}=-{3(N+2)\over2(N+8)^2}\log{4\over3}\,\epsilon^2\,+O(\epsilon^3)\,  ,
\eeqa
and the nonuniversal amplitude
\beq
A_R = 1+\epsilon^2 {3(N+2)\over 8(N+8)^2} \left(f(0)-4\gamma_E \log{4\over3}
 \right) +O(\epsilon^3)\, \, .
\eeq
For the {\it new universal function} $F_R(v)$ we find
\beq
F_R(v) = 1+\epsilon^2 {3(N+2)\over 8(N+8)^2} (f(v)-f(0)) + O(\epsilon^3)\; ,
\label{eqGR}
\eeq
A plot of the quantity $f(v)-f(0)$ 
(defined in the Appendix \ref{a1} Eq. (\ref{fvd})), 
that completely characterizes the 
out-of-equilibrium corrections to the mean-field behavior up to the second 
order in the $\epsilon$-expansion, is reported in Fig.~\ref{fv}.
Due to the small prefactor ($\epsilon^2/72$ for the Ising model, $N=1$), 
it might be very hard to detect these corrections in numerical and 
experimental works, 
as it happens for the corrections to the mean-field behavior of the 
static~\cite{PV-r} and equilibrium dynamics~\cite{2pteqdin} 
two-point functions.

\section{Two-loop fluctuation-dissipation ratio}
\label{sec4}

In this Section we evaluate the FDR up to the order $\epsilon^2$. 
We do not compute the full two-point correlation function $C(t,s)$, 
since only $\dpar_s C(t,s)$ is required to determine the FDR. 
This derivative may be computed by using the following diagrammatic identity.

\begin{figure}[t]
\centerline{\epsfig{file=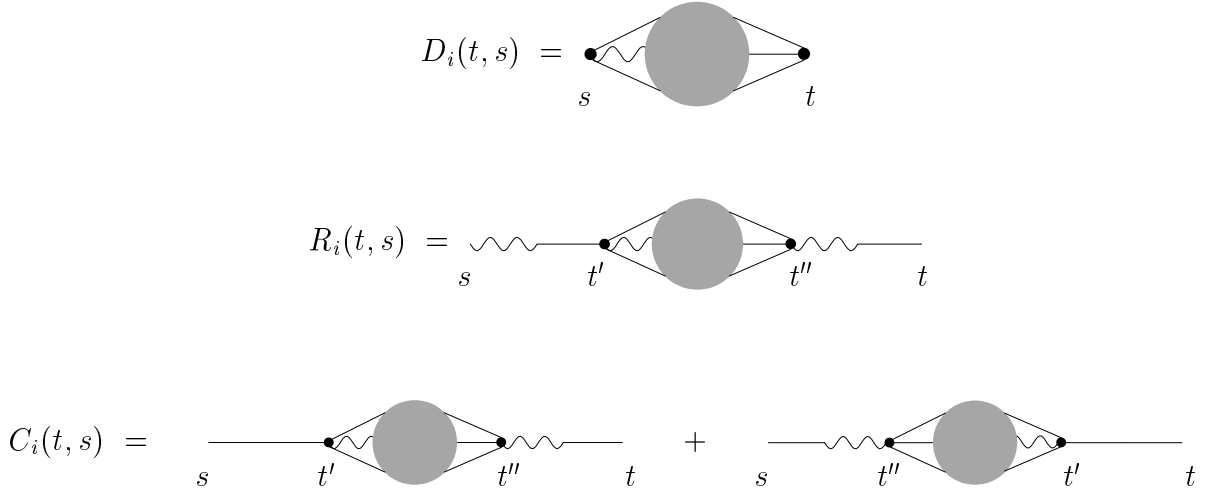}} 
\caption{Diagrammatic trick.}\label{fromRtoC}
\end{figure}

Each amputated diagram $D_i(t,s)$ (with label $i$) contributing to the 
response function, 
also contributes to
the correlation one in two diagrams, as graphically illustrated in Fig.
\ref{fromRtoC}.
Taking into account the explicit form of the propagators 
(see Eqs.~(\ref{Rgaux}) and (\ref{Cgaux})) for $q^2=0$
and causality (which also implies that $D_i(t,s)\propto\theta(t-s)$ apart 
from contact terms) it is easy to find that
\beq
\dpar_s C_i(t,s)= 2 R_i(t,s)+2 \int_0^\infty \d t' \, t' D_i(t',s),
\label{trick}
\eeq
where $C_i(t,s)$ is the contribution of this diagram to the correlation 
function, $R_i(t,s)$ the contribution to the response function,
and $D_i(t',s)$ the common amputated part.

\begin{figure}[b]
\centerline{\epsfig{file=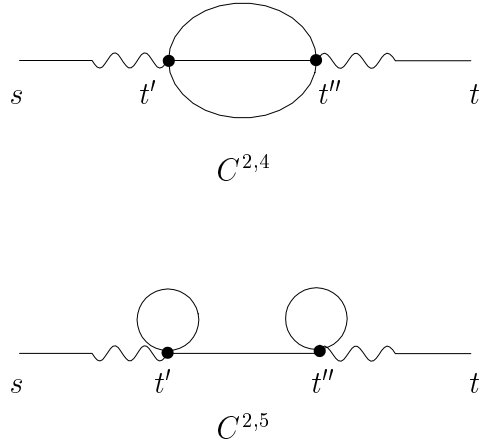}} 
\caption{Diagrams contributing only to the correlation function.
}\label{CnonR}
\end{figure}

Relation~(\ref{trick}) is nothing but a particular case of a  
relation following an algebraic identity for the functional integral, i.~e.
\beq
0 = \int [\d\varphi\d\tilde\varphi\d\varphi_0] \frac{\delta}{\delta\tilde\varphi({\bf x},s)}\left\{ \varphi({\bf x}',t) e^{-S[\varphi,\tilde\varphi]-H_0[\varphi_0]}\right\}\ ,
\eeq
with $t>s>0$. At criticality (i.e.~$r_0=0$, using dimensional 
regularization) we get in momentum space
\beq
(\partial_s -  {\bf q}^2)\langle\varphi(-{\bf q},t)\varphi({\bf q},s) \rangle = 2  \langle \varphi(-{\bf q},t)\tilde\varphi({\bf q},s) \rangle - \frac{g_0}{3!}\langle\varphi(-{\bf q},t)\varphi^3({\bf q},s)\rangle \ ,
\eeq
which, in the limit ${\bf q}^2 \rightarrow 0$, is diagrammatically expressed 
by Eq.~(\ref{trick}) as far as common amputated contributions to response and correlation functions are concerned.

Diagrams contributing to the correlation function, but not to the response 
one do exist. They have to be computed without taking advantage 
of this identity. At two-loop order there are two of them, 
as in Fig. \ref{CnonR}.

Summing the six contributions to the correlation function we 
finally arrive to the expression:
\beqa
{\dpar_s C_B(t,s)\over2}&=& R(t,s)-g_0 {N+2\over6}(\dpar C)^{1,1}_e+g_0^2 \left\{
\left({N+2\over6}\right)^2 (\dpar C)^{2,1}_e+{(N+2)^2\over18} (\dpar C)^{2,2}_e
\right. \nonumber \\&&\left.
+{N+2\over6} (\dpar C)^{2,3}_e+
{1\over2}\left[ {N+2\over18} (\dpar C)^{2,4}
+\left({N+2\over6}\right)^2 (\dpar C)^{2,5} \right]
\right\}+O(g_0^3)\, .
\eeqa

Considering the explicit expression for the diagrams given in the 
Appendix \ref{a2} one obtains the derivative of the bare 
correlation function.
This bare quantity is renormalized using equations (\ref{rencou}), 
(\ref{renfie}) and 
\beq
\Omega=Z_{\Omega} \Omega_B \qquad \mbox{ with }
\qquad Z_\Omega=\left({Z_{\varphi}\over Z_{\tilde{\varphi}}}\right)^{1/2}
\ ,
\eeq
so that, taking into account the $\Omega$ we set equal to $1$ in 
the previous relations,  
\beq
\partial_s C(t,s) = Z_{\Omega} Z_\varphi^{-1} \,\partial_s C_B(t,s) = (Z_\varphi Z_{\tilde\varphi})^{-1/2}\,\partial_s C_B(t,s) \ .
\eeq
The expression of $\dpar_s C(t,s)$ in terms 
of the renormalized coupling has a multiplicative 
redefinition of its amplitude at the first order in $\tilde{g}$.
Considering the fixed point value for $\tilde{g}$ (cf. Eq.~(\ref{gstar})) 
one finally obtains
\beqa
{\dpar_s C(t,s)\over2}&=&
\left[1 +\epsilon{N+2\over 4(N+8)}+\epsilon^2 
{ 3 (N+2)(3N+14) \over 4 (N+8)^3}\right]\nonumber\\
&&\times\left\{1+\epsilon {N+2\over 4(N+8)}\log {t\over s} +
{\epsilon^2 \over4}\left[
{6(N+2)\over (N+8)^2}\left({N+3\over N+8}+\log2\right) \log{t\over s}+
\right.\right.\nonumber\\&&\left.\left.
{(N+2)^2\over 8(N+8)^2}\log^2 {t\over s}
 -{6(N+2)\over (N+8)^2}\log {4\over3}\log(t-s)\right]\right\}
\nonumber\\
&&\times\Bigg\{1 +\epsilon^2 {N+2 \over (N+8)^2}\left[
{ 3  \over 4 } \log {4\over3} \log{t-s\over t+s}
-\gamma_E {3\over2} \log {4\over3}
+ \Phi(s/t) +{3\over8} f(s/t) + {N+2\over8}\right.\nonumber\\
&&\left.+{3\over2}\left(1-\log{4\over3}\right) \log 2
-{3\over4}\left(1+\log{4\over3}\right)+
{3\over 8}\log^2{4\over3}+{3\over4}\Li(1/4)
\right]\Bigg\} + O(\epsilon^3)\;,
\eeqa
where the function $f(v)$ and $\Phi(v)$ are defined in Eqs. (\ref{fvd}) and 
(\ref{phiv}) respectively, and $\Li$ is the dilogarithmic function whose
standard definition is recalled in Eq. (\ref{li2}).
Note that also for $\dpar_s C(t,s)$ all the nonscaling terms cancel out when 
the coupling constant is set equal to its fixed point value.
This result agrees with the scaling form in momentum space 
(analogous to Eq.~(\ref{dCscalform}))
\beq
\dpar_s C(t,s)= A_{\dpar C} (t-s)^a (t/s)^\theta F_{\dpar C} (s/t),
\eeq
with the same $a$ and $\theta$ as given in the previous section and a 
new universal scaling function $F_{\dpar C} (v)$ given by
\beq
F_{{\dpar C}} (v)=
1 +\epsilon^2 {3(N+2) \over 8(N+8)^2}\left[
2 \log {4\over3} \log{1-v\over 1+v}
+{8\over 3} (\Phi(v) -\Phi(0))+ f(v)-f(0) \right] + O(\epsilon^3)\, .
\label{eqGdC}
\eeq
A plot of the loop corrections in the above expression 
(apart from the factor ${3(N+2) \over 8(N+8)^2}$ appearing also in $F_R(v)$) 
is shown in Fig.~\ref{fv}. As already noticed for $F_R(v)$, effective 
corrections to mean-field behavior are quantitatively very small for 
$F_{{\dpar C}} (v)$.

Taking the long time limit (according to Eq. (\ref{Xq2})) of both the 
correlation and response functions one obtains 
the limit of the critical \fd ratio we are interested in:
\beq
{({\cal X}_{{\bf q}=0}^\infty)^{-1}\over2}=
1+ {N+2\over 4(N+8)} \epsilon+
\epsilon^2 {N+2\over(N+8)^2}\left[
{N+2\over8} +{3(3N+14)\over4 (N+8)}+ c
\right]+O(\epsilon^3)\, ,
\label{Xqbis}
\eeq
with 
\beqa
c&=& -\frac{3}{4} + {3\over4}\log2(2+11\log2-3\log3)-\frac{23}{8}\log^23
\nonumber\\&&
+{3\over2}\Li(1/4)-{21\over4}\Li(1/3)+{21\over8}\Li(3/4)-{1\over8}\Li(8/9) = 
-0.0415\dots\; .
\eeqa
We note that
the contribution of $c$ to the FDR is quite small. 
For example, with $N=1$ the sum of the first two 
terms in brackets is $\sim 1.8$, which is about $45$ times larger than $c$.

\section{Conclusions and Discussions}
\label{sec5}

In this work we studied the off-equilibrium properties of the  purely 
dissipative relaxational dynamics of an $N$-vector model in the framework of 
field theoretical $\epsilon$-expansion. 
The results presented here extend those of Ref.~\cite{cg-02}.
The scaling forms for the zero-momentum response function and for the 
derivative
with respect to the waiting time of the two-time correlation function read
\beqa
R(t,s)&=&A_R (t-s)^a (t/s)^\theta F_R(s/t)\, ,\\
\dpar_s C(t,s)&=& A_{\dpar C} (t-s)^a (t/s)^\theta F_{\dpar C} (s/t).
\eeqa 
The universal functions $F_R(s/t)$ and $F_{\dpar C} (s/t)$ are given in 
Eqs. (\ref{eqGR}) and (\ref{eqGdC}) respectively.
In both cases the corrections to the Gaussian value $1$ is of order 
$\epsilon^2$.
In principle these corrections should be detectable
in computer and experimental works, but being quantitatively very small, they 
are hard to observe.
We would remark that this fact does not mean that aging effects in these models
are weak compared with the analogous phenomena in glassy systems. 
In fact aging manifests itself in the full scaling forms (e.g. 
$\theta\neq 0$) and in the violation of \fd theorem, i.e. in $X^\infty\neq 1$
in a quantitative way.
 
We note that the $R(t,s)$ we found agrees with the general RG form, but
at first sight it is not compatible with the Fourier transform of 
Eq. (\ref{henkelpred}). This na\"{\i}ve comparison should be done very 
carefully because it involves a Fourier integral which could be divergent.
The analysis of the full $q$-dependence of $R_{\bf q}(t,s)$ may give some
insight into this problem.
This dependence has been already carried out up to $O(\epsilon)$ \cite{cg-02}, 
but it is very hard to determine it up to two loops. 
In other dynamical universality classes this discrepancy already arises at 
 $O(\epsilon)$. The computation of the full $q$-dependence in these cases 
seems to be simpler and may provide some useful hints \cite{prep}.

We computed the FDR ${\cal X}_{\bf q=0}$ for general $N$, 
cf. Eq.~(\ref{Xqbis}). 
As shown in Ref.~\cite{cg-02} this quantity for zero momentum has the 
same long-time limit as the standard FDR $X^\infty$.
Using this equality we may
compare our result with those presented in the literature.

In the limit $N\rightarrow \infty$, Eq.~(\ref{Xqbis}) reduces to 
$X^{\infty}=1/2-\epsilon/8 -\epsilon^2/32+O(\epsilon^3)$, in agreement with
the exact result for the spherical model $X^{\infty}=1-2/d$ \cite{gl-00}.

\begin{figure}[t]
\centerline{\epsfig{file=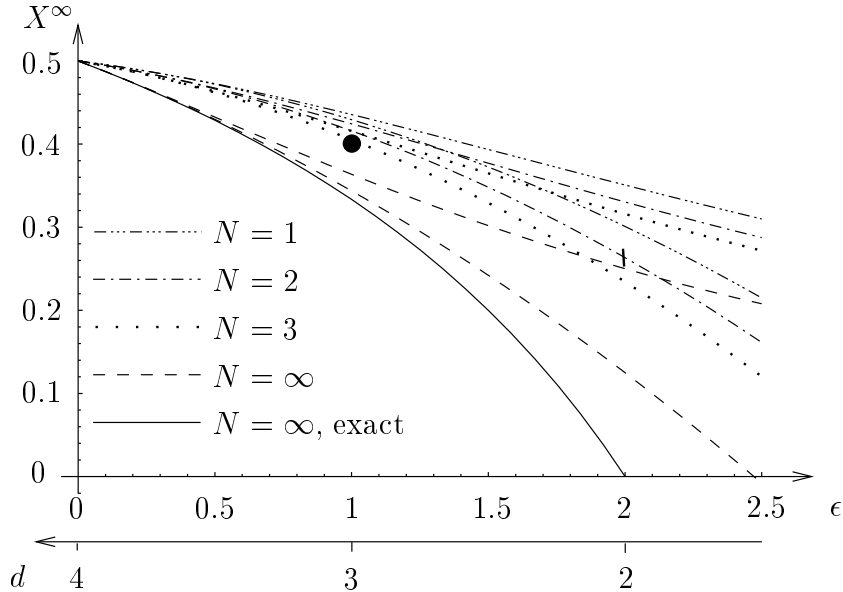}} 
\caption{$X^\infty$ as a function of the dimensionality $d=4-\epsilon$ for
several $N$. For each $N$ the upper curve is the $[2,0]$ Pad\'e approximant 
and the lower one the $[0,2]$. The exact result for $N=\infty$ is
reported as a solid line. The numerical Monte Carlo values for 
the Ising Model in two and three dimensions are also indicated (for the 
latter, there is no indication about the error).}
\label{figfixN}
\end{figure}

\begin{figure}[t]
\centerline{\epsfig{file=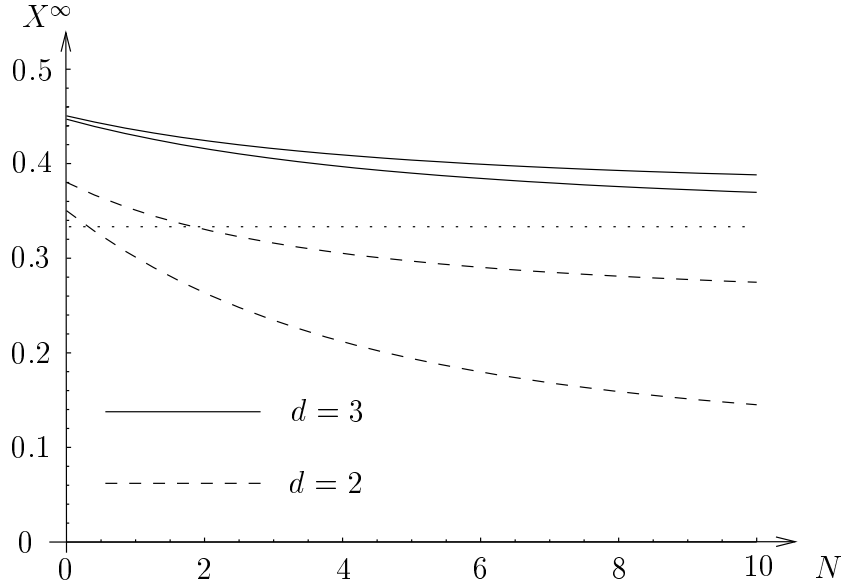}} 
\caption{$N$ dependence of $X^{\infty}$ for $d=2,\,3$.
The upper curve is the $[2,0]$ Pad\'e approximant 
and the lower one the $[0,2]$.
The dotted  line is the exact result for $N=\infty$ 
in $d=3$~($X^\infty=1/3$)
}\label{figfixd}
\end{figure}

The formula for general $N$ (cf. Eq. (\ref{Xqbis})) allows us to
make quantitative prediction for a large class of systems.
In Fig. \ref{figfixN} we report the dependence of $X^{\infty}$
on the dimensionality at fixed $N$, while in Fig. \ref{figfixd} we show the 
dependence on $N$ at fixed $d=4-\epsilon$.
For each model we report two values: one is obtained by direct summation
(Pad\'e approximant $[2,0]$) and the other by ``inverse'' summation 
(Pad\'e approximant $[0,2]$). We do not show the $[1,1]$ approximant,
since it has a pole in the range of $\epsilon$ we are interested in.
From these figures some general trends may be understood:
\begin{itemize}
\item 
decreasing the dimensionality, $X^{\infty}$ always decreases, at least
up to $\epsilon=2$ (for the one-dimensional Ising model the value
$X^{\infty}=1/2$ is expected \cite{gl-00});
\item increasing $N$, $X^{\infty}$ decreases, approaching in a quite 
fast way the exact result for the spherical model;
\item for $N=\infty$ the curve of the $[0,2]$ approximant reproduce better
the exact result in any dimension with respect to the $[2,0]$ approximant.
\end{itemize}

The last point suggests the use of the $[0,2]$ value as estimate of  
$X^{\infty}$, also for physical $N$. We quote as {\em indicative} error
the difference between the two approximants.
Using this procedure, we obtain $X^{\infty}=0.429(6)$ for the 
three-dimensional $N=1$ model,
compared to $\simeq 0.46$ found at one-loop~\cite{cg-02}, in very good
agreement with the Monte Carlo simulation value 
$X^{\infty}\simeq 0.40$ for the three-dimensional Ising Model~\cite{gl-00}
with nonconservative (heat-bath Glauber) dynamics.
Considering $\epsilon =2$ one obtains 
$X^{\infty}=0.30(5)$ for $N=1$, improving the one loop estimate 
$\simeq 0.42$ in the right direction towards the Monte Carlo result
$X^{\infty}=0.26(1)$ for the two-dimensional Ising Model with 
Glauber dynamics~\cite{gl-00}.

Using our results we can give predictions of $X^{\infty}$ for systems 
that have not yet been analyzed by numerical simulations. We estimate 
$X^{\infty}=0.416(8)$ for the three-dimensional $XY$ model and 
$X^{\infty}=0.405(10)$ for the three-dimensional Heisenberg model.
These predictions may be tested by numerical simulations extending the
results quoted in \cite{gl-00}.

There are also several open questions that need further 
investigations. For example a ``rigorous'' proof of the fact
that the FDR is exactly $1$ for all modes with $q^2\neq 0$ 
(somehow related to the presence of a mass gap) has not yet been given.
Then one might ask how these theoretical results 
(scaling forms, relaxing modes etc.) change if one changes mesoscopic dynamics
(e.g. with conserved quantities), or when more complex static 
Hamiltonians are considered, e.g. those with disorder or frustration, 
or when different initial conditions (e.g. with long-range correlations)
are considered.
We will consider in forthcoming works the $O(N)$ model with Model B and C 
dynamics \cite{prep} and the purely 
dissipative relaxation of the Ising model with quenched random impurities
\cite{cg-02b}.

\section*{Acknowledgments}
The authors are grateful to M.~Henkel for useful correspondence and 
comments and to S.~Caracciolo, A.~Pelissetto, and E.~Vicari for a critical 
reading of the manuscript.

\appendix

\section{Connected diagrams for the response function}
\label{a1}

The four diagrams contributing to the response function up to the two-loop
order are reported in Fig. \ref{figR}.
The one-loop diagram was already discussed in Ref. \cite{cg-02}.
The expression obtained there for the critical bubble (i.e. for the 1PI part 
of the diagram) is
\beq
B_c(t)= \int \! {\d^dq \over (2 \pi)^d} \; C_q^0(t,t)=-
{1\over d/2-1}{(2 t)^{1-d/2} \over (4 \pi)^{d/2}}=
- N_d {\Gamma(d/2-1)\over 2^{d/2}} t^{1-d/2}.
\eeq
Thus the full connected 1-loop diagram for the response function is given by
\beqa
R^{1,1}(t,s)&=&\int_s^t \d t' B_c(t')= 
- N_d {\Gamma(d/2-1)\over 2^{d/2} (2-d/2)} (t^{2-d/2}-s^{2-d/2})\nonumber\\
&=&- N_d {1\over4} \left[ \log{t\over s}+ {\epsilon\over2} 
\left((\gamma_E+\log2+\log t)
\log{t\over s}-{1\over2} \log^2{t\over s}\right)\right]+O(\epsilon^2) \ .
\label{A2}
\eeqa

From these one-loop expressions, it is quite simple to compute the two-loop
integrals $R^{2,1}$ and $R^{2,2}$ of Fig.~\ref{figR}.
Indeed the two-loop critical bubble~(the 1PI part of $R^{2,1}$) can be
computed in terms of $B_c(t)$ as
\beq
B_{c2}(t)= \int  {\d^dq_1 \over (2 \pi)^d} \int_0^\infty \d t'
\; B_c(t') R^0_{q_1}(t,t') C^0_{q_1}(t,t')=
N_d^2 \frac{b(d)}{4-d} t^{3-d}\, ,
\eeq
where
\beq
b(d)= {\Gamma^2(d/2-1) \over 2^{d-1}} 
\left[1- {(4-d) \Gamma^2(2-d/2) \over 2 \Gamma(4-d) }\right]=
-{1\over8}\ [1+\epsilon (\log 2+\gamma_E)+O(\epsilon^2)].
\eeq
By means of  this expression, we compute $R^{2,1}$ taking into account the
external legs with $q=0$
\beq
R^{2,1}(s,t)=\int_s^t \d t' B_{2c}(t')=
N_d^2 b(d) {t^{4-d}-s^{4-d}\over(4-d)^2}\, ,
\eeq
that near four dimensions has the following series expansion 
\beq
R^{2,1}(s,t)={N_d^2\over8}\left[
-\log{t\over s}\left({1\over\epsilon}+\log t+\log2 +\gamma_E\right)
+{1\over2}\log^2{t\over s}\right]+O(\epsilon).
\eeq

The computation of $R^{2,2}$ is simple once  the expressions for $R^{1,1}$ and
$B_c(t)$ are known. Indeed, from Eq.~(\ref{A2}), it is obtained:
\beq
R^{2,2}(s,t)=\int_s^t \d t' R^{1,1} (t,t') B_c(t')=
N_d^2 {\Gamma^2(d/2-1)\over 2^{d-1}}
\left[{t^{2-d/2}-s^{2-d/2}\over{4-d}}\right]^2\; ,
\eeq
that is, expanding in $\epsilon$,
\beq
R^{2,2}(s,t)=N_d^2{1\over32}\log^2{t\over s}+O(\epsilon)\; .
\eeq

The last diagram $R^{2,3}$ is more difficult to be worked out and it requires
 a long calculation which main steps are described in what follows.
First of all we evaluate its 1PI contribution called $O_1(t,s)$
\beqa
O_1(t,s)&=& \int {\d^d q_1 \over (2 \pi)^d} \int {\d^d q_2 \over (2 \pi)^d}
C^0_{q_1}(t,s) C^0_{q_2} (t,s) R^0_{q_1+q_2} (t,s) \nonumber\\
&=&\theta (t-s) \int {\d^d q_1 \over (2 \pi)^d}  {\d^d q_2 \over (2 \pi)^d}
{1\over q_1^2  q_2^2}
(e^{-q_1^2(t-s)}-e^{-q_1^2(t+s)}) (e^{-q_2^2(t-s)}-e^{-q_2^2(t+s)}) 
e^{-(q_1+q_2)^2 (t-s)}\nonumber\\
&=& \theta (t-s) \left[ 
(t-s)^{2-d} J_d (1,1) + (t+s)^{2-d} J_d \left(1,{t-s \over t+s}\right)
- 2 (t-s)^{2-d} J_d \left( {t+s \over t-s},1 \right) \right]\; ,
\eeqa
with
\beq
J_d(a,b)= \int {\d^d q_1 \over (2 \pi)^d} \int {\d^d q_2 \over (2 \pi)^d}
e^{-q_1^2-a q_2^2 -b(q_1+q_2)^2}=
N_d^2 [(1+b)(a+b)]^{1-{d\over2}} F_d \left({4b^2 \over  (b+1)(a+b)}\right)\; ,
\eeq
and
\beq
F_d(x)= {\Gamma(d/2)\over4}\int_0^\infty \d s s^{d/2-2} e^{s x/4} \Gamma(0,s)=
\frac{\Gamma(d/2) \Gamma (d/2-1)}{2(d-2)} \,_2F_1
\left({d\over 2} -1,{d\over 2} -1,{d\over 2},{x\over 4}\right)\; .
\eeq
In particular, for our calculations, we are interested in the limits
\beqa
F_4(x)&=& -{\log (1-x/4) \over x},\label{Ffour}\\
F_{4-\epsilon}(1)&=& 
\log {4\over 3}+\epsilon\left[\left(\gamma_E-{1\over 2}\right)\log {4\over 3}- 
{1\over4}\log^2 {4\over 3}+{1\over2}\Li \left({1\over 4}\right)\right] 
+O(\epsilon^2)\, ,\\
F_d(0)&=& {\Gamma^2 (d/2-1) \over 4}.
\eeqa
Here $\Li(z)$ is the standard dilogarithm, defined as
\beq
\Li(z)=\sum_{k=1}^{\infty}{z^k\over k^2}\; .
\label{li2}
\eeq
The final expression for $O_1(t,s)$, in generic dimension, is
\beq
O_1(t,s)={N_d^2 \, \theta(t-s) \over 2^{d-2}}
\left[ F_d(1) (t-s)^{2-d}+t^{2-d} F_d\left(\left({t-s \over t}\right)^2\right)
-2 (t(t-s))^{1-d/2} F_d\left({t-s \over t}\right)\right]\; .
\eeq
The full connected diagram  $R^{2,3}(s,t)$ is thus given 
by the following expression
\beq
R^{2,3}(s,t)=\int_s^t \d t''\int_{t''}^t \d t' O_1(t',t'')= N_d^2
(A_1(s,t)+A_2(s,t)-2 A_3(s,t)),
\label{A16}
\eeq
where
\beqa
A_1(s,t) &=& 2^{2-d} F_d(1) \displaystyle{{(t-s)^{4-d}\over (4-d)(3-d)}}\;,\\
A_2(s,t) &=&2^{2-d} t^{4-d}\int_{s/t}^1 \d y \,y^{3-d}\int_y^1 \d z \,z^{d-4} F_d((1-z)^2) =2^{2-d} t^{4-d} I_1 (s/t)\;,\\
A_3(s,t) &=&2^{2-d} t^{4-d}\int_{s/t}^1 \d y \, y^{3-d}\int_y^1 \d z \, z^{d-4} (1-z)^{1-d/2} F_d(1-z) = 2^{2-d} t^{4-d} I_2 (s/t)\;.
\eeqa
The evaluation of the two functions $I_1(v)$ and $I_2(v)$ is rather cumbersome 
but algebraically trivial. After some calculations one gets
\beqa
I_1(v)&=&{\Gamma^2(d/2-1)\over 4}[\log v(8\log 2-6\log3)+f_1(v)+O(\epsilon)],\\
I_2(v)&=&{\Gamma^2(d/2-1)\over 4}\left[-{2\over \epsilon} \log v-\log^2v-
(1-6\log 2+3\log 3)\log v +f_2(v)+O(\epsilon)\right],
\eeqa
where $f_i(v)$ are given by
\beqa
f_1(v)/4&=&\log v\int_0^v \d z \,F_4((1-z)^2)+\int_v^1 \d z \, \log z \,F_4((1-z)^2),\\
f_2(v)/4&=&\log v\int_0^v \d z \, {F_4(1-z)-1\over 1-z}+
\int_v^1 \d z \,{\log z\over 1-z}[F_4(1-z)-1]+
\int_v^1 \d z{\log (1-z) \over z},
\eeqa
and in particular these are regular functions in the limit $v\rightarrow 0$ 
\beqa
f_1(0)&=&\log^2 2 +\log^2{8\over3}+3 \Li (1/4)-4\Li(2/3)\,,\\
f_2(0)&=&-{\pi^2\over 6}+{3\over2} \log^2{4\over3}-\Li (1/4).
\eeqa
Inserting all these contributions in Eq.~(\ref{A16}), we get
\beqa
{4 R^{2,3}(s,t)\over N_d^2}&=&
 -{1\over\epsilon} \left(\log {4\over3}+\log{t\over s}\right)-
\log {4\over3}(\log(t-s)+\gamma_E)\nonumber\\
&&-({1\over2}+\gamma_E+\log t)\log{t\over s}
+{1\over2}\log^2{t\over s}+{f(s/t)\over4}+O(\epsilon),
\label{eqR23}
\eeqa
with 
\beqa
f(v)&=&f_1(v) - 2f_2(v)-\log{4\over3} (2+\log12) - 2\Li(1/4),\label{fvd}\\
f(0)&=&\frac{\pi^2}{3} - 2\log{4\over3} +3\log^22 -\log^2{8\over3}  
   + 3\Li(1/4)-4\Li(2/3)\nonumber\\
&=& 0.663707 \dots.
\eeqa

\section{Connected diagrams for the FDR}
\label{a2}

In this Appendix we evaluate the rest of the diagrams required for the 
computation of the FDR.
We do not evaluate the full integral for the correlation function, since
we make use of the trick explained in details in Section~\ref{sec4}.
For this reason we consider first those diagrams contributing also to
the response function and we evaluate only their extra-contributions (given by
$\int_0^\infty \d t'' t'' D_i(t'',s)$ in Eq.~(\ref{trick}) and denoted with 
the subscript ``$e$'' in what follows) to the derivative of the correlation 
function.
For the first three diagrams these contributions are very simple:
\beqa
(\dpar C)_e^{1,1}&=& s B_c(s)= N_d\left[-{1\over4}-{\epsilon\over8}(\log s+\gamma_E+\log2)\right]+O(\epsilon^2)\, , \\
(\dpar C)_e^{2,1}&=& \int_0^s \d t'' t'' B_c(t'') B_c(s)= N_d^2 {\Gamma^2(d/2-1)\over2^d(3-d/2)} s^{4-d}={N_d^2\over 16}+O(\epsilon), \\
(\dpar C)_e^{2,2}&=& s B_{c2}(s)=-N_d^2 {1\over8} ({1\over\epsilon}+\gamma_E+\log2+\log s)
+O(\epsilon)\, .
\eeqa

The fourth contribution is less simple
\beqa
(\dpar C)_e^{2,3}&=&\int_0^s \d t'' t'' O_1(t'',s) = N_d^2 2^{2-d}s^{4-d}
\Bigg[ {F_d(1)\over(4-d)(3-d)} \nonumber \\
&&+ \int_0^1 \d z\, z F_d ((1-z)^2)- 
2 \int_0^1 \d z\, z(1-z)^{1-d/2} F_d(1-z)\Bigg].
\eeqa
Using now the explicit form for $F_4(x)$ given in Eq.~(\ref{Ffour}), one 
obtains
\beqa
{4(\dpar C)_e^{2,3}\over N_d^2}&=&-\left({1\over\epsilon}+\log s +\gamma_E+{1\over2}\right)
(\log{4\over3}+1)+\Li(1/4)+
\log{4\over3}\left({1\over4}\log{4\over3}-\log2\right) +O(\epsilon)\, .
\eeqa

The diagrams whose amputated part does not contribute also to the response 
function are shown in Fig. \ref{CnonR}. 
The sunset-type diagram $(\dpar C)^{2,4}$ is quite difficult, 
thus we first compute its 1PI part $O_2(t,s)$.
Introducing $q_3=q_1+q_2$, this contribution is given by (for $t>s$)
\beqa
O_2(t,s)&=& \int {\d^d q_1 \over (2 \pi)^d} \int {\d^d q_2 \over (2 \pi)^d}
C^0_{q_1}(t,s) C^0_{q_2} (t,s) C^0_{q_3} (t,s) \nonumber\\
&=& \int {\d^d q_1 \over (2 \pi)^d} \int {\d^d q_2 \over (2 \pi)^d}
\prod_i {1\over q_i^2} (e^{-q_i^2(t-s)}-e^{-q_i^2(t+s)})\nonumber\\
&=& N_d^2\left[\Delta^{3-d} K_d(1)+3 \sigma^{3-d} K_d(\Delta/\sigma)-
3\Delta^{3-d} K_d(\sigma/\Delta)-\sigma^{3-d} K_d(1)\right],
\eeqa
with $\Delta=t-s$, $\sigma=t+s$, and
\beqa
K_d(x)&=& {1\over N_d^2}\int {\d^d q_1 \over (2 \pi)^d} \int {\d^d q_2 \over (2 \pi)^d}
{1\over q_1^2}{1\over q_2^2}{1\over (q_1+q_2)^2}
e^{-q_1^2-q_2^2-x (q_1+q_2)^2}\nonumber\\
&=&
{\Gamma(d/2-1)\Gamma(d/2)\over 4} \int_x^\infty 
{\d u \over(1+u)^{d-2}}\int_0^{1}\d v v^{d/2-2}
\left[1-{v u^2\over (1+u)^2}\right]^{1-d/2}.
\eeqa
In the following we are interested in the limits
\beqa
K_4(x)&=&{1\over2}\log {2(1+x) \over 1+2x}
-{1\over4x}\log{1+2x\over (1+x)^2}\;,\\
K_{4-\epsilon}(1)&=& 
{3\over4} \log{4\over 3}+ {\epsilon\over4} 
\left[3\log{4\over3} \left({1\over2}+\gamma_E\right)
+{1\over 4}\log^2 3+{\Li(8/9)\over 2} \right]
+O(\epsilon^2) \; .
\eeqa
Introducing these results in the expression for the connected diagram 
\beq
(\dpar C)^{2,4}=\int_0^t \d t' O_2(t',s)=\int_0^s O_2(s,t')+\int_s^t O_2(t',s),
\eeq
one finds
\beqa
{(\dpar C)^{2,4}\over N_d^2}&=&{K_d(1)\over4-d}\left(2s^{4-d}+(t-s)^{4-d}-(t+s)^{4-d}\right)\nonumber\\
&&+3\int_0^s \d t'\left[(s+t')^{3-d} K_d\left({s-t'\over s+t'}\right)-
(s-t')^{3-d} K_d\left({s+t'\over s-t'}\right)\right]\nonumber\\
&&+3\int_s^t \d t' \left[(s+t')^{3-d} K_d\left({t'-s\over t'+s}\right)-
(t'-s)^{3-d} K_d\left({s+t'\over t'-s}\right)\right]\nonumber\\
&=&{K_d(1)\over4-d}\left(2s^{4-d}+(t-s)^{4-d}-(t+s)^{4-d}\right)\nonumber\\
&&+3(2s)^{4-d}\left[\int_0^1 \d y (1+y)^{d-5}[K_d(y)-y^{3-d}K_d(1/y)]
\right.\nonumber\\
&&\left.
+\int_0^{t-s\over t+s} \d y (1-y)^{d-5}[K_d(y)-y^{3-d}K_d(1/y)]\right]
\nonumber\\
&=&{3\over2}\log{4\over3}\left({1\over\epsilon}+\log s +{1\over2}\log{t-s\over t+s}+\gamma_E\right)
+\Phi(s/t)+ O(\epsilon)\, ,
\eeqa
where
\beqa
\Phi(v)&=&2 K'_4(1)-{3\over2}\gamma_E\log{4\over3}+\nonumber\\ 
&& + 3\left[ \int_0^1 {\d y \over 1+y}\left(K_4(y)-{1\over y} K_4(1/y)\right)
+ \int_0^{1-v\over1+v} {\d y\over1-y}\left(K_4(y)-{1\over y} K_4(1/y)\right)
\right] .\label{phiv}
\eeqa
In particular we are interested in the limit $v\rightarrow0$, given by
\beqa
\Phi(0)&=&
2K'_4(1) + 6\int_0^1 {\d y \over 1-y^2}\left[K_4(y)-{1\over y} K_4
\left({1\over y}\right)\right]-{3\over2}\gamma_E\log{4\over3}\nonumber\\
&=& {3\over4}\log{4\over3}+{39\over4}\log^2 2-{9\over4}\log2+\log3-
{13\over4}\log^2 3-{21\over4}\Li(1/3) \nonumber\\
&&+{21\over8} \Li(3/4)-{1\over8} \Li(8/9) = -0.24889\dots
\eeqa

Now the only diagram left is $C^{2,5}$ of Fig.~\ref{CnonR}. 
It is given by
\beqa
C^{2,5}(t,s)&=&\int \d t' \d t'' R^0_{q=0}(t,t'')B_c(t'')C^0_{q=0}(t'',t')B_c(t') R^0_{q=0}(s,t)\nonumber\\
&=&N_d^2 {\Gamma^2(d/2-1)\over 2^{d-1} (3-d/2)(2-d/2)}
\left[t^{2-d/2}-{s^{2-d/2}\over 5-d}\right] s^{3-d/2}\, .
\eeqa
Its derivative with respect to $s$, near four dimension, is
\beq
(\dpar C)^{2,5}=\dpar_s C^{2,5}(t,s)= N_d^2 {1\over 8}\left[\log {t\over s}+1\right]
+O(\epsilon)\; .
\eeq



\begin{references}
\bibitem{footnote0} We are not interested here in the problem of phase 
ordering dynamics (see Ref.~\cite{bray} for a review), so we will not consider the case of a quench
to $T<T_c$.
\end{references}
\end{document}